\documentclass[useAMS,usenatbib]{mn2e}

\usepackage{graphicx}
\usepackage{mathrsfs}
\usepackage{natbib}
\usepackage{amssymb}
\usepackage{amsmath}
\usepackage{mathtools}
\bibpunct[,]{(}{)}{;}{a}{}{,}

\def\sun{{\odot}}
\newcommand{\Msun}{\ensuremath{M_\odot}}
\newcommand{\Ye}{\ensuremath{Y_{\rm e}}}

\newcommand{\kms}{\ensuremath{{\rm km~s}^{-1}}}
\newcommand{\tns}{\ensuremath{t_{\rm ns}}}
\newcommand{\erad}{\ensuremath{\dot{\epsilon}_{\rm rad}(t)}}

\title[Kilonovae light curves from disk winds]{Kilonova Light Curves from the Disk Wind Outflows of Compact Object Mergers}
\author[Kasen, Fern\'andez, \& Metzger]{Daniel Kasen$^{1,2,3}$, Rodrigo Fern\'andez$^{1,2}$, Brian D. Metzger$^{4}$\\
\\
$^1$ Department of Astronomy and Theoretical Astrophysics Center, University of California, Berkeley, CA 94720, USA.\\
$^2$ Department of Physics, University of California, Berkeley, CA 94720, USA.\\
$^3$ Nuclear Science Division, Lawrence Berkeley National Laboratory, Berkeley, CA 94720, USA.\\
$^4$ Columbia Astrophysics Laboratory, Columbia University, New York, NY 10027, USA.\\
}

\begin{document}

\date{Submitted to MNRAS}
\pagerange{\pageref{firstpage}--\pageref{lastpage}} 
\pubyear{2013}
\maketitle
\label{firstpage}

\begin{abstract}
We study the radioactively-powered transients produced by accretion disk
winds following a compact object merger.  Starting with the
outflows generated in two-dimensional hydrodynamical disk models, we use
wavelength-dependent radiative transfer calculations to generate
synthetic light curves and spectra.  We show that the brightness and
color of the resulting {\it kilonova} transients carry information about the
merger physics.  In the regions of the wind where neutrino
irradiation raises the electron fraction to $\Ye \gtrsim 0.25$, r-process
nucleosynthesis halts before producing  high-opacity, complex ions (the
lanthanides).  The kilonova light
curves thus show two distinct components: a brief ($\sim 2$ day) blue
optical transient produced in the outer lanthanide-free ejecta,
and a longer ($\sim 10$ day) infrared transient produced in the inner,
lanthanide line-blanketed region.  Mergers producing a longer-lived neutron star, or a more
rapidly spinning black hole, have stronger neutrino irradiation, generate more lanthanide-free ejecta, and
 are optically brighter and bluer.  At least
some optical emission is produced in all disk wind models, which should
enhance the detectability of electromagnetic counterparts to
gravitational wave sources.  However, the presence of even a small
amount ($10^{-4}~M_\odot$) of overlying, neutron-rich dynamical ejecta
will act as a ``lanthanide-curtain", obscuring the optical wind
emission from certain viewing angles.
Because the disk outflows have moderate velocities ($\sim 10,000~\kms$),
numerous resolved line features are discernible in the
 spectra,  distinguishing disk winds from fast-moving dynamical
ejecta, and offering a potential diagnostic of the detailed composition
of freshly produced r-process material.
\end{abstract}


\maketitle

\section{Introduction}

\begin{table*}
%
\caption{\label{t:models}Disk wind model properties and summary of radiative transfer results. The first two
columns from the left show model name and lifetime of the HMNS, respectively. The
following five columns show properties of the homologous ejecta: mass with
with $Y_e < 0.25$, mass with $Y_e>0.25$, kinetic energy, mean velocity of material with $Y_e<0.25$,
and mean velocity of material with $Y_e > 0.25$. The last three columns give the peak luminosities
in the blue ($3500-5000$~\AA), red ($5000-7000$~\AA), and infrared ($1-3\mu$m) bands, respectively.}
\begin{tabular}{lcccccccccc}
{Model}&
{$t_{\rm ns}$} &
{$M_{Y_e < 0.25}$} &
{$M_{Y_e > 0.25}$} &
{KE} &
{$\bar{v}_{Y_e < 0.25}$} &
{$\bar{v}_{Y_e> 0.25}$} &
{$\nu L_\nu({\rm B})$} &
{$\nu L_\nu({\rm R})$} &
{$\nu L_\nu({\rm IR})$}\\
{ } & {(ms)} & {($M_\sun$)} & {($M_\sun$)} & {ergs} & {(km s$^{-1}$)} & {(km s$^{-1}$)} &  {(ergs/sec)} &  {(ergs/sec)} & {(ergs/sec)} \\
\hline
t000    &  0      &    $1.5 \times 10^{-3}$ & $6.9 \times 10^{-5}$  &  $1.6 \times 10^{48}$  & 8,927 & 18,583 &
$4.3 \times 10^{40}$ &  $2.1 \times 10^{40}$ &  $8.5 \times 10^{39}$ \\
t030    &  30      &    $2.7 \times 10^{-3}$ & $5.4 \times 10^{-4}$ & $9.9 \times 10^{48}$ & 7,424 & 29,518  &
$5.8 \times 10^{40}$ &  $6.0 \times 10^{40}$ &  $1.0 \times 10^{40}$ \\
t100    & 100      &   $7.3 \times 10^{-4}$ & $5.3 \times 10^{-3}$ & $2.5 \times 10^{49}$ & 9,805 & 15,007 &
$6.6 \times 10^{40}$ &  $1.3 \times 10^{41}$ &  $5.7 \times 10^{39}$ 
\\
t300   &  300  & - & $1.5 \times 10^{-2}$  & $6.9 \times 10^{49}$  & - &  16,432  &
$ 8.5 \times 10^{40}$ & $2.4 \times 10^{41}$ & $1.2 \times 10^{40}$ 
\\
tInf    & $\infty$ &    - &  $2.9 \times 10^{-2}$  & $1.9 \times 10^{50}$ & - &  21,419  &
$2.7 \times 10^{41}$ &  $3.9 \times 10^{41}$ &  $2.0 \times 10^{40}$ 
\\
\noalign{\smallskip} 
a0.8    &   0      &    $4.9 \times 10^{-3}$ &  $8.4 \times 10^{-4}$ & $9.9 \times 10^{48}$ & 9,996 & 21,012 &
$1.3 \times 10^{41}$ &  $7.5 \times 10^{40}$ &  $2.1 \times 10^{40}$ 
\\
\hline
\end{tabular}
\end{table*}

The ejection of radioactive material during, or immediately following,
the merger of two neutron stars (or a neutron star and a black hole)
can give rise to an optical/infrared transient similar to, but dimmer
and briefer than, an ordinary supernova (\citealt{Li&Paczynski98};
\citealt{Kulkarni05}; \citealt{Metzger+10}; \citealt{Roberts+11};
\citealt{Barnes&Kasen13}; \citealt{Piran+13};
\citealt{Tanaka&Hotokezaka13}; \citealt{Grossman+14};
\citealt{Tanaka+14}).  These transients, called {\it kilonovae}, are
promising electromagnetic counterparts to gravitational wave sources
(\citealt{Metzger&Berger12,Nissanke+13}), and may be diagnostic of the
sites of heavy element nucleosynthesis (\citealt{Lattimer&Schramm74};
\citealt{Freiburghaus+99}; \citealt{Goriely+11}; \citealt{Piran+14}).

Most studies of kilonovae have focused on material that becomes
unbound during the merger itself.  Simulations find that, on a
dynamical timescale of $\sim$ milliseconds, around $10^{-4} - 10^{-2}
M_\odot$ of material may be flung out from the tips of the tidal tails
or squeezed out from the contact interface between the two coalescing
stars (e.g., \citealt{Rosswog05}; \citealt{Duez+10};
\citealt{Hotokezaka+13}; \citealt{Bauswein+13}).  This dynamical
ejecta is initially very neutron rich, with an electron fraction $\Ye
= n_p/(n_n + n_p) \lesssim 0.1$, where $n_p$ and $n_n$ are the number
density of protons and neutrons, respectively.  Neutrinos emitted by
the hot merger remnant can irradiate the outflowing material, perhaps
raising the $\Ye > 0.1$ along the polar rotation axis
\citep{Wanajo+14}.

Recently, more attention has focused on the possibility that a
comparable or greater amount of mass may be expelled subsequent to the
merger, as material in a rotationally supported disk accretes onto the
central remnant on a viscous timescale ($\sim 1$~s) (e.g.,
\citealt{Metzger+08}; \citealt{Beloborodov08}; \citealt{Metzger+09}).
Viscous and nuclear heating during accretion can drive
winds that unbind a significant fraction of the disk
(\citealt{Lee+09}; \citealt{Fernandez&Metzger13}; \citealt{Just+14}).
Neutrino irradiation during this phase will change the composition of the outflows.
If the central remnant is a black hole (BH), neutrinos emitted from
the inner region of the disk will raise the mean electron fraction of the wind
to $\Ye \sim 0.2$ (\citealt{Surman+06}; \citealt{Metzger&Fernandez14}).  
If the BH is rapidly spinning, this neutrino
irradiation is enhanced, leading to higher values of $Y_e$
(\citealt{Just+14}; \citealt{Fernandez+14}).  If the central remnant
survives for some period of time as a hyper-massive neutron star
(HMNS) before collapsing to a BH, the neutrino irradiation will be
even stronger, and the electron fraction can be raised to yet higher
values, $\Ye \sim 0.3$ (\citealt{Metzger&Fernandez14,Perego+14}).

The value of \Ye\ in large part determines the final composition of
the ejecta, and has a dramatic effect on the kilonova light curves.
Unbound material will assemble into heavy elements via rapid neutron
captures, or the r-process.  The line opacity of some heavy elements
 differs significantly from that of ordinary astrophysical mixtures.
In particular, for  species with open $f-$shell valence electron
configurations, namely the lanthanides (atomic number $Z = 58-70$) and
actinides ($Z = 89-103$), the high atomic
complexity leads to extremely high opacities \citep{Kasen+13}.  The high lanthanide opacity
 leads to a dimmer, longer-duration kilonova light
curve with emission peaking at infrared, rather than optical,
wavelengths \citep{Barnes&Kasen13}.\footnote{An important exception occurs if the outermost ejecta expands sufficiently rapidly for neutrons to avoid capture into nuclei.  The high radioactive heating rate of free neutrons relative to r-process nuclei may power bright ultraviolet/blue emission on a timescale of hours following the merger, despite the presence of lanthanides (\citealt{Metzger+14}).}

For ejecta with low electron fraction, r-process nucleosynthesis
proceeds all the way up to the third r-process peak, and a significant
fraction of lanthanides, and perhaps actinides \citep{Eichler+14}, are produced.  For
higher electron fraction, nucleosynthesis stops at the second
r-process peak, and the ejecta remains lanthanide-free.  The color and
brightness of kilonova light curves are therefore sensitive markers of
the electron fraction of the ejecta, and can be used to gain insight
into the physics of compact object mergers.

The predicted colors of disk winds will  also impact 
 strategies for finding electromagnetic
 counterparts to  the gravitational wave sources
detected by the advanced LIGO/VIRGO experiments.  Most
existing and upcoming wide-field surveys observe in optical bands;
infrared surveys are less common and typically have much lower
sensitivity and a smaller field of view.  The presence of optical
emission from a lanthanide-free disk wind would therefore enhance
the prospect of finding a kilonova.

Here we present radiative transfer calculations of kilonova light
curves that allow us to connect the observables
to the merger physics, e.g., the relative mass of
dynamical to wind ejecta, the lifetime of a HMNS, or the spin of a
central BH.  We begin with the 2-dimensional hydrodynamical
simulations from \citet{Metzger&Fernandez14} and \citet{Fernandez+14}
which model the post-merger evolution (\S\ref{sec:hydro}) and predict
the distribution and electron fraction of unbound wind ejecta
(\S\ref{sec:ejecta}).  In addition, we cary out r-process reaction network
calculations for parameterized wind outflows to estimate the critical
electron fraction above which the  ejecta is lanthanide free
(\S\ref{sec:nucleo}).  We then use a multi-dimensional radiative
transfer code to generate synthetic kilonova spectra and light curves
as a function of viewing angle (\S\ref{sec:radtrans}).  We study
several different models in order to explore the effects of the
lifetime of a HMNS (\S\ref{sec:HMNS}), the spin of a BH
(\S\ref{sec:bhspin}), and the presence of overlying dynamical ejecta
(\S\ref{sec:dyn}).  We compare our wind  light curves to
observations of potential kilonova candidates (\S\ref{sec:obs}), and
reflect on what implications our results have for detecting and
interpreting kilonovae in the future (\S\ref{sec:conclude}).

\section{Properties of Disk Winds}
\label{s:initial_conditions}

\subsection{Hydrodynamical  Method}
\label{sec:hydro}

We consider six hydrodynamical disk wind simulations from
\cite{Metzger&Fernandez14} and one from \citet{Fernandez+14}, as shown
in Table~\ref{t:models}.  All calculations begin with an equilibrium
torus of mass $0.03~\Msun$ surrounding a central remnant. Models
differ in the amount of time that the neutron star survives before
collapse to a black hole, with values of $t_{\rm ns} = \{0, 30,
100,300\}$~ms. For one model (tInf), the neutron star was assumed to
survive indefinitely.  All models assume a non-rotating black hole
with the exception of model $a0.8$, which considers the case of the
prompt formation of a black hole with spin parameter $a = 0.8$.  We
constructed several additional ejecta models, discussed in
\S\ref{sec:dyn}, by superimposing an outer shell or torus
of dynamical ejecta upon the wind model t100.

The hydrodynamical simulations use the  \textsc{FLASH3.2} code
\citep{dubey2009}, with modifications that enable modeling the viscous
evolution of merger remnant accretion disks
\citep{F12,Fernandez&Metzger13,FM12}. The code includes a Helmholtz
equation of state \citep{timmes2000} with abundances in nuclear
statistical equilibrium, charged-current neutrino rates including
emission from a hypermassive neutron star and disk self-irradiation
via a neutrino leakage scheme, and viscous angular momentum transport
via an $\alpha$-viscosity prescription
\citep{shakura1973}. Approximate general relativistic effects are
included via the pseudo-Newtonian potential of \citet{paczynsky1980}
for non-spinning BHs and HMNSs, and the potential of
\citet{artemova1996} for the spinning BH case.  All models are evolved
until $3000$ orbits at the initial density maximum, or $\sim 10$~s.

To follow the unbound ejecta all the way to free-expansion, we use a
two-step process. First, in the original disk simulations we record
all material leaving a sphere of radius $10^9$~cm, centered on the
central remnant.  This information is then used as an inner boundary
condition for a calculation on a new, larger computational domain of
radius $10^{14}$~cm. The hydrodynamics are then carried out without
any source terms other than gravity and the equation of
state\footnote{Viscous and neutrino source terms operate on timescales
  slower than the expansion time at these radii.}. The ambient density
is reduced to $10^{-8}$~g~cm$^{-3}$ to prevent deceleration of the
wind.  By the time the ejecta reaches $10^{12}$~cm, the material is
very nearly homologous, i.e., the kinetic energy dominates and the
velocity is proportional to radius.  From this time on, the dynamics
can be extrapolated analytically.

\subsection{Ejecta Properties}
\label{sec:ejecta}

\begin{figure}
\includegraphics*[width=\columnwidth]{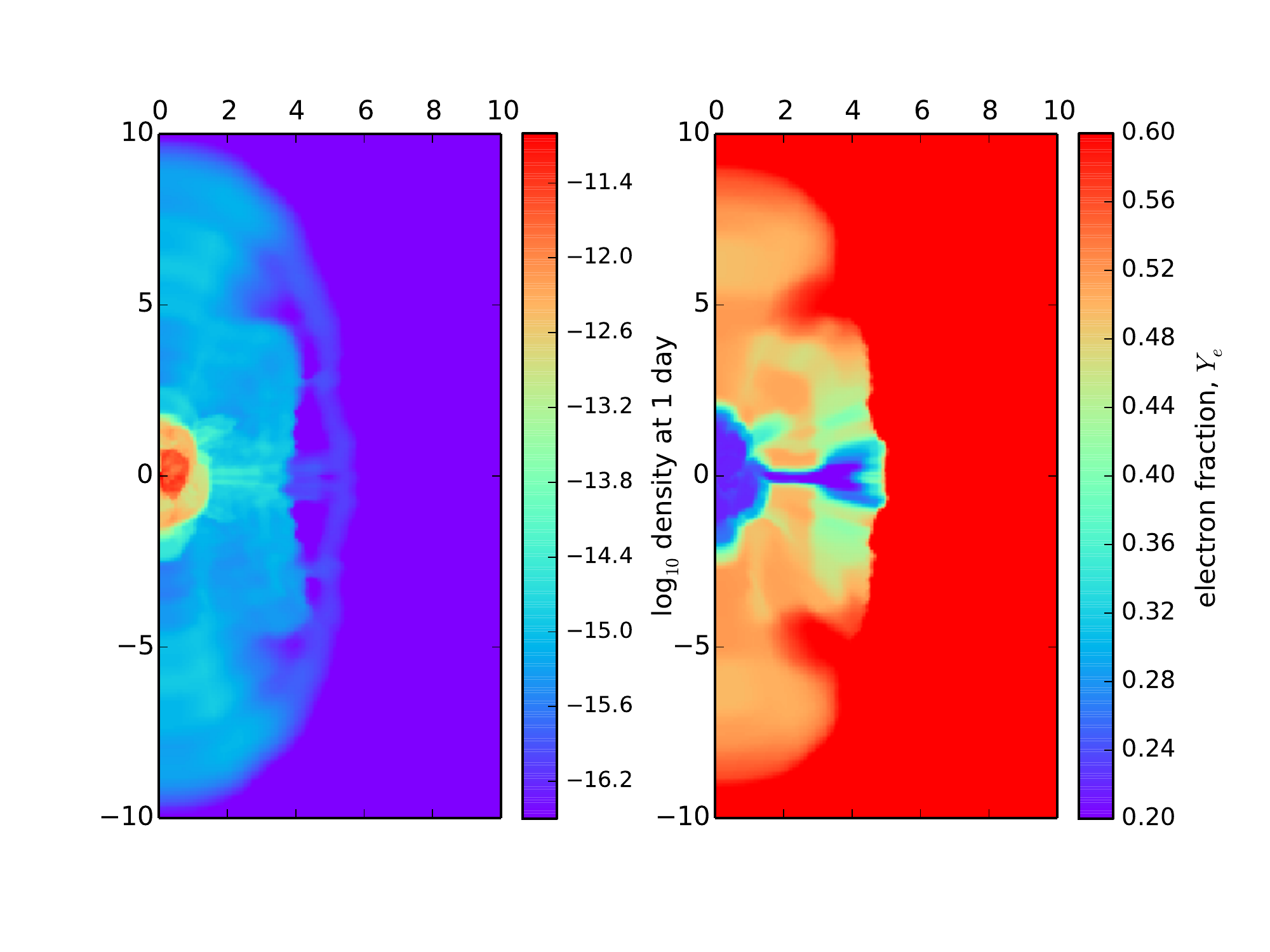}
\caption{
\label{fig:model}
Density (left) and electron-fraction (right) structure of the ejecta from model t030 at 1 day after  merger.  In this phase, the ejecta is in homologous expansion, with velocity proportional to radius.  The axes are given in velocity
coordinates with units of $10^9~{\rm cm~s^{-1}}$. 
}
\end{figure}

\begin{figure}
\includegraphics*[width=\columnwidth]{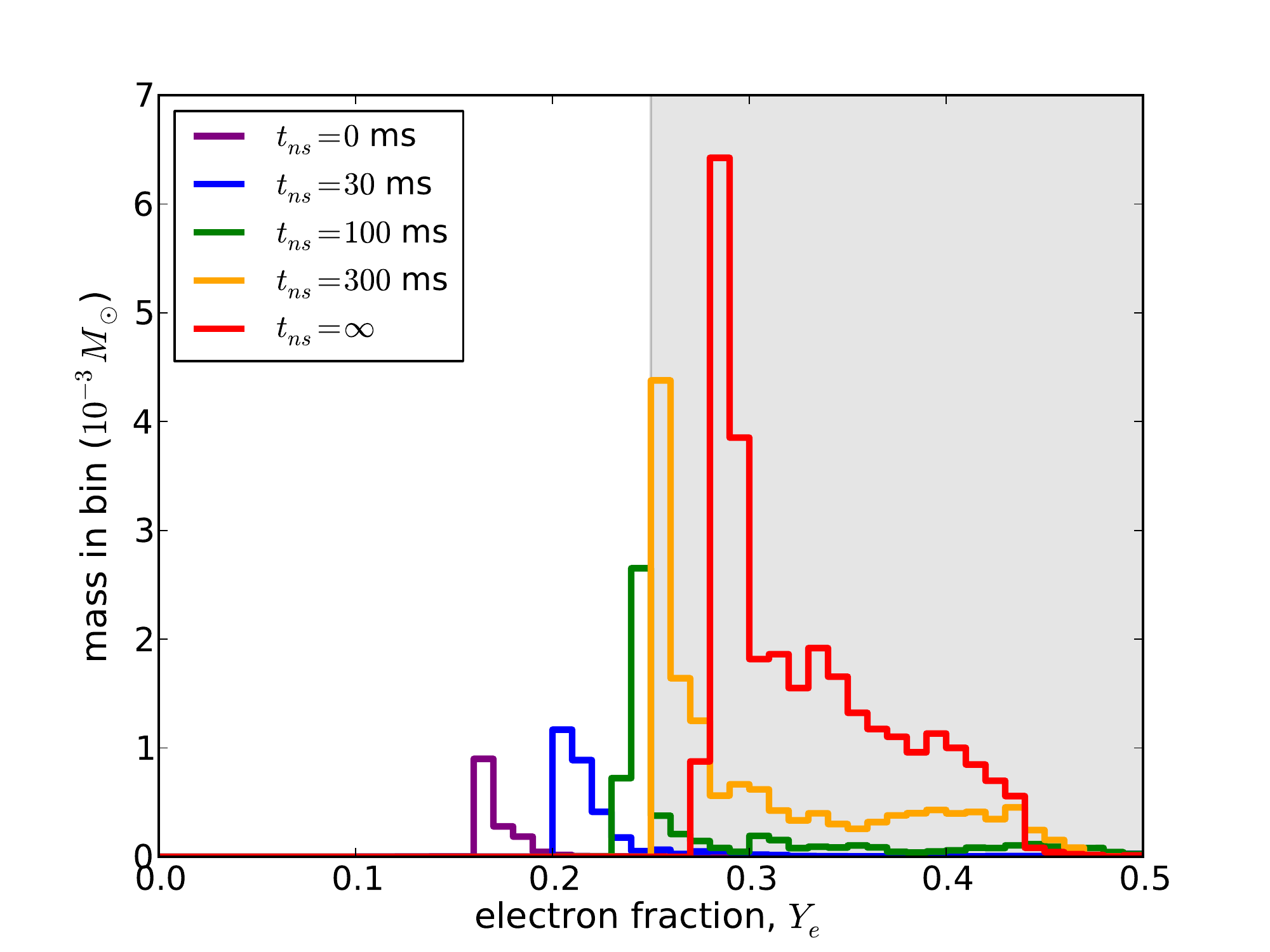}
\caption{
The amount of mass ejected with different values of the electron fraction, for disk wind
models of different  neutron star lifetimes, $t_{\rm ns}$.  A longer-lived neutron star leads to a
larger total ejected mass  and a
higher mean electron fraction.  The shaded area shows the region $Y_e > 0.25$ where  
the ejecta is likely lanthanide-free.
\label{fig:ye_mass}
}
\end{figure}

Figure~\ref{fig:model} shows the density and compositional structure
of a representative model, t030, in the homologous phase.  The ejecta
is approximately spherical, with the bulk of the material moving at
speeds of $\sim 10,000$~\kms, or $0.03c$.  Such expansion velocities
are substantially lower than that of the dynamical ejecta, which moves
at $0.1-0.3c$.

The outer layers of wind ejecta typically have a higher electron
fraction than the inner regions, as the outer material is ejected at
earlier times when the neutrino irradiation is higher due to the
presence of a neutron star or a higher accretion rate onto the black
hole.  In model t030, a low \Ye\ plume is also seen along the
equator. This feature forms out of the motion of the highly-irradiated
component of the wind, which originates in regions of small radius and
high altitude above the disk midplane, and wraps around the back of
the disk with near north-south symmetry \citep{Fernandez+14}.
The mass in this plume, however, is very small compared to the
spherical core.

Figure~\ref{fig:ye_mass} shows, for each model, a histogram of the
amount of mass ejected with various values of \Ye.  Increasing the
lifetime of a HMNS has two main consequences,  First, a larger
total amount of mass is ejected, due to higher
neutrino heating and the presence of a hard boundary at the HMNS,
which keeps disk material from being swallowed below the event horizon of a
BH.  Second, the mean value of \Ye\ increases with \tns\ due to the
greater level of neutrino irradiation from both the HMNS and the disk.

\subsection{Nucleosynthesis and Lanthanide Fraction}
\label{sec:nucleo}

\begin{figure}
\includegraphics*[width=\columnwidth]{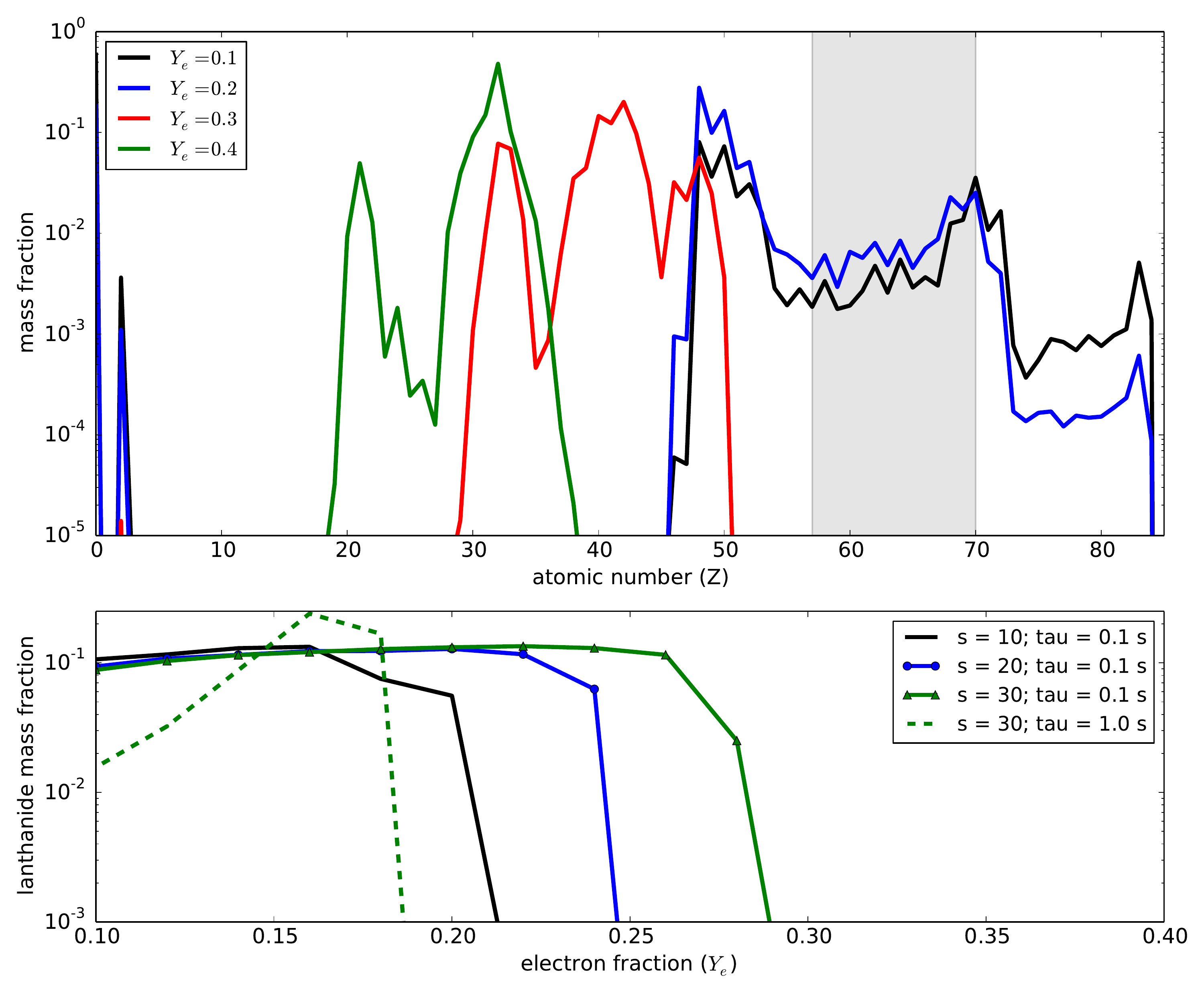}
\caption{
Nucleosynthesis results for parameterized wind models. {\it Top panel:}
Final mass fractions of the wind for an expansion time
$t = 0.1$~s, entropy = $20 k_{\rm B}/$baryon and varying values of electron fraction, $Y_e$.  
For higher values of $Y_e$, the r-process  fails to produce lanthanides (shaded grey region).
 {\it Bottom panel:}  Integrated  mass fraction of all lanthanides as a function of electron fraction
 assuming different values of entropy and expansion time.  For conditions typical of disk wind ejecta ($s = 20 k_{\rm B}$/baryon, $t = 0.1$~s), no lanthanides are produced if $Y_e \gtrsim 0.25$.
\label{fig:torch}
}
\end{figure}

Knowledge of the compositional structure of the wind ejecta is needed
to predict the resulting kilonova light curves.  A full calculation of
the ejecta abundances would require detailed nuclear-reaction network
post-processing of thermodynamic
trajectories along the wind.
Here we approximate the composition by a one-to-one mapping from the
electron fraction.  This mapping is obtained by evolving parameterized
trajectories with the nuclear reaction network code Torch
\citep{Timmes+99}.  Trajectories begin with abundances in nuclear
statistical equilibrium at a temperature $T = 5\times 10^9$~K and
entropy $s\simeq 20$~k$_{\rm B}$/baryon. The density decays
exponentially in time, with expansion time $t_{\rm exp}=100$~ms and
the entropy is held constant, assuming that radiation dominates the
pressure. The chosen values of entropy and expansion time correspond
to mass-flux weighted averages from disk wind simulations at the point
where the average temperature is $5\times 10^9$~K
\citep{Fernandez&Metzger13}.  Torch lacks a treatment of fission,
which is an important physical effect in low-\Ye\ outflows, but should
not be significant around the critical \Ye\ at which lanthanides first
appear.

The top panel of Figure~\ref{fig:torch} shows the final mass fraction
distributions for a few sample wind trajectories of different \Ye.  We
find that the lanthanide mass fraction decreases sharply above $\Ye
\approx 0.25$.  For electron fractions above this value, the neutron
density is too low for the $r$-process to proceed past the long
beta-decay lifetimes associated with the $N=82$ closed shell nuclei,
and nucleosynthesis halts at atomic numbers of $Z \approx 58$. For
higher electron fractions, however, the flow is able to move past this
point and proceeds rapidly to the next closed shell at $N=126$.

The bottom panel of Figure~\ref{fig:torch} shows that the transition
to a lanthanide-free composition is fairly abrupt in \Ye, and not
overly sensitive to the value of the entropy chosen, so long as it is
a few times $10$~k$_{\rm B}$/baryon.  We therefore consider $\Ye \approx
0.25$ as a critical value for distinguishing between the presence or
absence of lanthanides.

\section{Light Curves and Spectra}

\begin{figure*}
\includegraphics[width=6.5in]{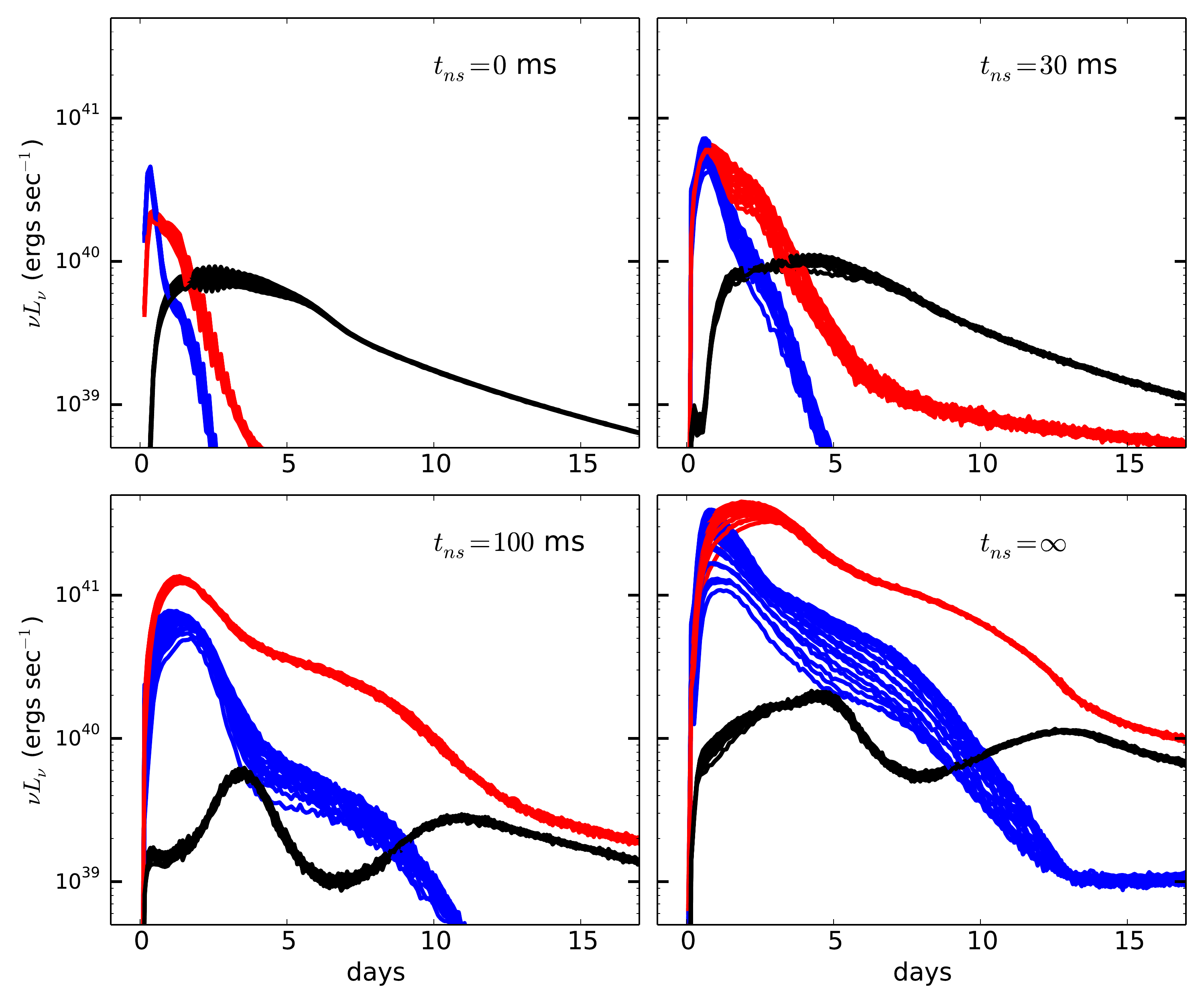}
\caption{
Synthetic light curves of models t000, t030,  t100, and tInf.  Blue lines denote blue optical emission (averaged over $3500-5000$~\AA), red lines denote red optical emission ($5000-7000$~\AA), and black lines
infrared emission ($1-3~\mu$m).  Lines are overplotted for 20 different viewing angles, equally spaced in $\cos \theta$.  A small amount of Monte Carlo noise is apparent in the calculations.
\label{fig:br_lc}
}
\end{figure*}

\subsection{Radiative Transfer Method}
\label{sec:radtrans}

We input the homologous wind profiles from the hydrodynamical
simulations of \S\ref{sec:ejecta} into the multi-dimensional radiative
transfer code Sedona \citep{Kasen+06} to calculate synthetic light
curves and spectra.  The calculation setup was similar to that
discussed in \citet{Kasen+13} and \citet{Barnes&Kasen13}, and assumed
local thermodynamic equilibrium for the atomic level
populations and ionization state.

As the line data of  high-Z elements is still incomplete, we used
an approximate method to construct effective opacities of r-process
mixtures.  The opacity of all d-shell species was calculated using the
atomic data for iron \citep{Kurucz_Bell_1995}, while the opacity of
all f-shell species (the lanthanides) was calculated using the
detailed atomic structure calculations for neodymium ($Z = 60$)
\citep{Kasen+13}.  Atomic structure calculations have shown that this
proxy approach provides a reasonable approximation of the
pseudo-continuum opacities, although it fails to produce the proper
line features. We used the wind nucleosynthesis calculations of
\S\ref{sec:nucleo} to determine the abundance of lanthanides, and
assumed that the remainder of the ejecta included the 30 d-shell
species between $Z = 21$ and $Z = 80$.
 
For the radioactive heating rate, \erad, of the wind, we use the
results given in \cite{Roberts+11}, which we assume to be the same in all
regions of ejecta with $\Ye < 0.4$.   In reality, the heating rate will depend
on the local electron fraction.  \cite{Grossman+14} show that, on day timescales, the heating rate of
 $\Ye \approx 0.1-0.3$ ejecta can be a factor of $\sim 2$ greater than that of the $\Ye < 0.1$ outflows
calculated by \cite{Roberts+11}.  Our calculations may therefore  underestimate
the radioactive heating  and kilonova luminosities.   For $\Ye > 0.4$, the heating rate 
is lower \citep{Grossman+14} and so we
take $\erad = 0$ in these regions. Since most of the ejecta
have $Y_e < 0.4$, this choice has little impact on the final light
curves.

The Sedona calculations generate the spectral time series every 0.1~days
after merger, within a wavelength range of  $200-30000$~\AA, and from
20 different viewing angles equally spaced in the cosine of the polar
angle $ \theta$.  From the spectra, we constructed broadband light
curves by averaging the emission over three different wavelength
ranges: $3500-5000$~\AA\ (``blue"), $5000-7000$~\AA\ (``red") and $1-3
\mu$m (``infrared").  Table~1 gives the peak luminosity of all models
in each of these bands.

\subsection{Effect of Neutron Star Lifetime}
\label{sec:HMNS}

\begin{figure*}
\includegraphics*[width=6.8in]{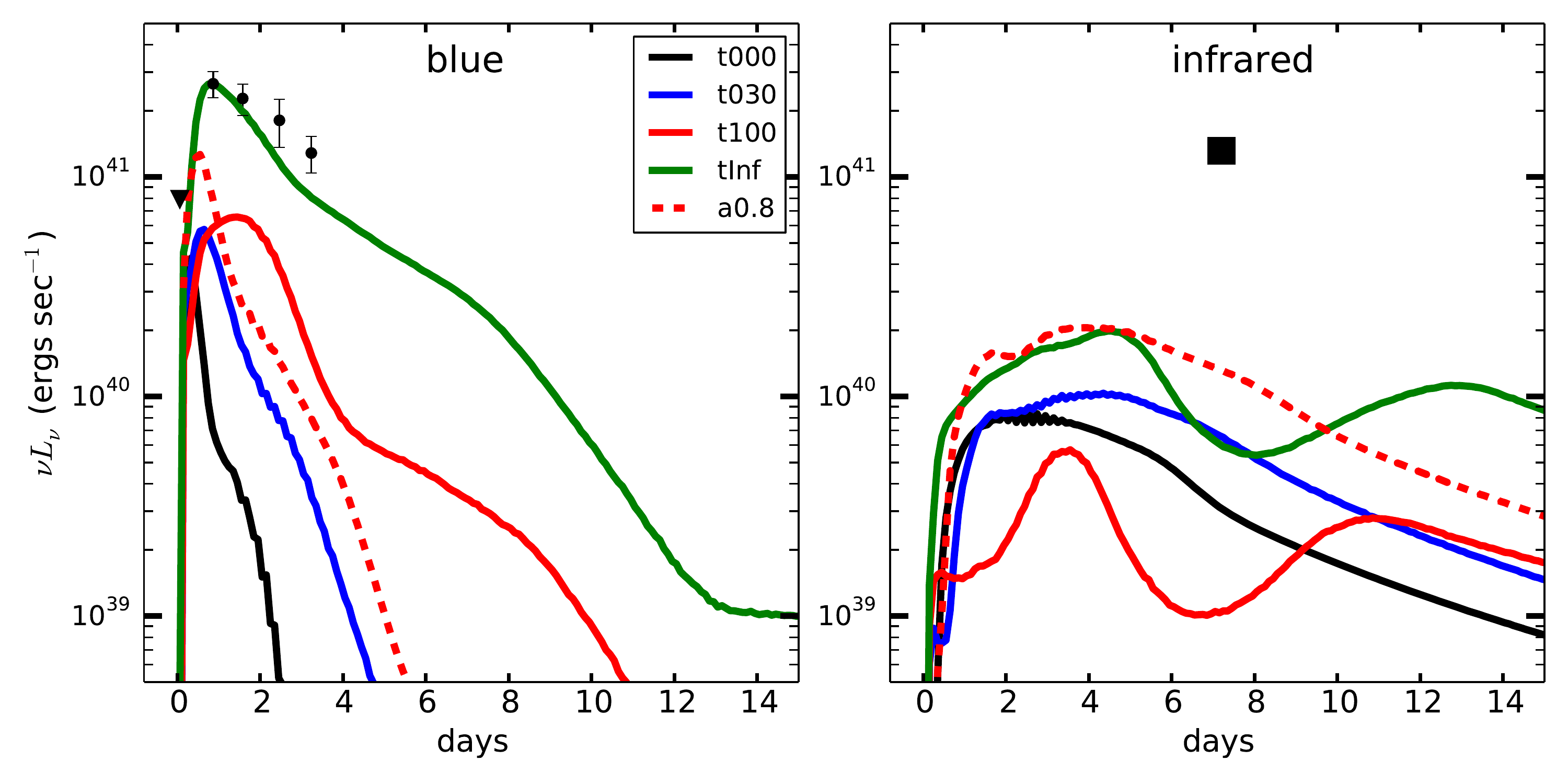}
\caption{
 {\it Left Panel:} Angle averaged synthetic light curves of various wind models at  optical blue wavelengths ($3500-5000$~\AA). The closed circles show r-band observations of the possible kilonova following GRB 080503 \citep{Perley+09}. The triangle symbol denotes an upper limit.  As the redshift of 080503 is unknown, we adopt a value $z = 0.25$ and neglect k-correction effects.  
{\it Right Panel:} Model light curves of the same models at  infrared wavelengths ($1-3 \mu$m).  The square shows the Hubble Space Telescope  observations of the possible kilonova associated with
GRB130603B \citep{Tanvir+13, Berger+13}.
\label{fig:allmods}
}
\end{figure*}

Figure~\ref{fig:br_lc} shows the predicted light curves, from all
viewing angles, of four models with different NS lifetimes.
Figure~\ref{fig:allmods} shows angle-averaged light curves for the
same models.  Because the wind outflows are roughly spherically
symmetric, the variation of the light curves with viewing angle is
generally small.  The exception is model tInf, where the ejecta have a
significant prolate elongation, and the blue light curve is a factor
of $\sim 3$ brighter for an equatorial viewing angle, at which the
projected surface area of the ejecta is maximal.

The light curves of model t000 -- which assumes prompt formation of a BH -- show two distinct components: a brief ($\sim
2$~day) blue optical transient, and a longer ($\sim 10$~day) infrared
transient. The infrared emission is generated within the central
ejecta, where the electron fraction is $\Ye < 0.25$ and heavy line
blanketing from the lanthanides suppresses the optical flux.  This
light curve is qualitatively similar to that found in the
parameterized models calculated by \cite{Barnes&Kasen13}.  The blue
component of the light curve is produced by the small amount of high
\Ye\ (lanthanide-free) material in the outer layers of ejecta.  The
production of even a small ($\sim 10^{-4}~M_\odot$) amount of
high-\Ye\ in the exterior ejecta therefore has important implications
for the detectability of kilonovae with optical facilities.

As the lifetime of the HMNS is increased, the optical light curves
become brighter.  This is because neutrino irradiation converts a
larger fraction of the wind to $\Ye > 0.25$, and a greater total mass
is ejected due to the hard boundary of the HMNS.  For the extreme case
of a stable NS (model tInf), the entire wind is lanthanide-free and
 the blue optical emission peaks at $\nu
F_\nu \approx 2.8 \times 10^{41}~{\rm ergs~s^{-1}}$, a factor of 10
brighter than that of the prompt BH model t000.  The optical light
curves roughly follow analytic expectations that the duration should
scale as $t \propto M^{-1/2}$ and the peak optical luminosity as $L
\propto M^{1/2}$ \citep{Metzger+10}, where $M$ is the ejected mass of
high \Ye\ material given in Table~1.

The dependence of the infrared brightness on \tns\ is non-monotonic.
The mass of low \Ye\ ejecta in model t030 ($\tns = 30$~ms) is greater
than that of model t000, and hence the infrared light curve brighter.
A turnover point, however, is reached around $\tns \approx 100$~ms, at
which point the neutrino irradiation is sufficient to convert nearly
the entire wind to high-\Ye, reducing the infrared emission.  The kilonova
colors thus correlate with the degree of neutrino irradiation.  For
$\tns \lesssim 30$~ms, the ratio $\nu L_\nu(B) / \nu L_\nu(IR) \approx
5$, whereas for $\tns > 30$~ms the color is much bluer, with $\nu
L_\nu(B) / \nu L_\nu(IR) \approx 10$.

The origin of the infrared emission in models with $\tns \gtrsim 100$~ms is
distinct  from those with $\tns < 100$~ms.  In the latter cases, the
infrared emission arises in low-\Ye, lanthanide-blanketed regions of
ejecta.  In the former, there is no low-\Ye\ ejecta, and the infrared
emission is simply the long wavelength tail of the thermal spectrum
that peaks in the optical.  Such infrared light curves display two
distinct maxima, separated by about 10 days.  The origin of the
secondary maximum is similar to that studied for Type Ia SN, and
results from an enhancement in infrared emissivity that occurs when
the ejecta transitions from doubly to singly ionized \citep{Kasen+06}.
The clear separation of the two maxima in these models may be an
artifact of our approximate opacity prescription, which uses iron
group atomic data as a proxy for all d-shell elements.  In reality, the
 change  in ionization state occurs at
a different time for different elements, depending on the ionization
potential.  For complex mixtures, this may have the effect of smearing the two peaks
together.

\begin{figure}
\includegraphics*[width=\columnwidth]{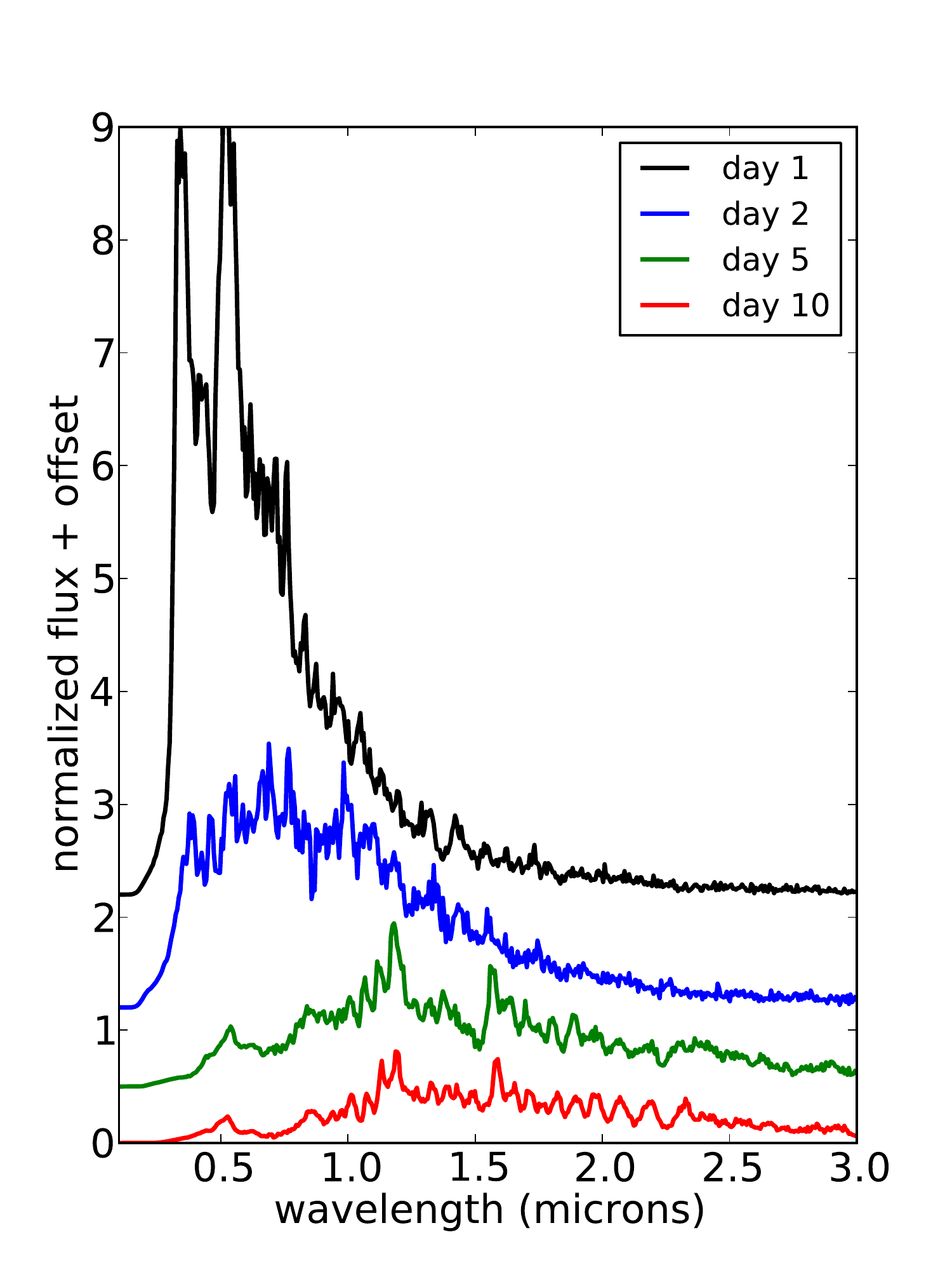}
\caption{Spectral time series of model t100.   The spectrum at early times ($t \lesssim 2$~days) is dominated by
optical emission from the lanthanide-free, outer layers of the wind.  At later times, the emission is dominated
by infrared emission from lanthanide-rich material in the central wind regions.
  Relatively narrow line features ($v \approx 5000~{\rm km~s^{-1}}$) 
are distinguishable however, due to the uncertainties in the atomic data, the positions of these features are not reliable.
\label{fig:spec_series}
}
\end{figure}

Figure~\ref{fig:spec_series} shows the spectra evolution of model t100
over 10 days.  The color of the continuum rapidly evolves from blue
emission produced in the outer high-\Ye\ layers of ejecta to infrared
emission produced in the inner, low-\Ye\ ejecta.  The spectra show
numerous line absorption features that, given the moderate ejecta
velocities ($5000-10,000~\kms$), are fairly well resolved.  This
differs from the spectra of dynamical ejecta, where the line features
are highly blended due to the fast  ($0.1-0.3$c)
ejecta velocities \citep{Kasen+13}.  However, given our approximate line opacities, the
position of individual lines can not be trusted, and as line data from
more species is added, line blending may become more prevalent.
Although more work is needed to make quantitative spectral
predictions, our results suggest that the relative slowness of the wind may be
discernible in the line features, providing a way to
distinguish a wind from dynamical ejecta.  The presence of resolved
lines also provides hope that the detailed composition of outflows
could be estimated from spectral analysis.

\subsection{Effect of Black Hole Spin}
\label{sec:bhspin}

We have also considered one model, a0.8, in which the central remnant
immediately became a rapidly spinning BH with Kerr parameter $a=0.8$.
The BH spin leads to a deeper potential well and a higher neutrino
irradiation from the inner regions of the disk.  The fraction of
high-\Ye\ material in the wind is therefore greater than it is in the
non-spinning prompt BH model t000.  The amount of high and
low-\Ye\ material ejected in model a.08 turns out to be comparable to
the case of a HMNS with lifetime $\tns = 30$~ms.  As seen in
Figure~\ref{fig:allmods}, the blue and infrared light curves of models
a0.8 and t030 are indeed quite similar.  This degeneracy represents a
challenge in using kilonova observations to diagnose the lifetime of a
HMNS.

\subsection{Effect of Dynamical Ejecta}
\label{sec:dyn}

Prior to the ejection of disk winds, neutron rich material may be
dynamically expelled in the merger, producing an overlying layer of
rapidly expanding, presumably low-\Ye\ material. This dynamical ejecta
may act as a ``lanthanide curtain", obscuring our view of the optical
emission originating with the wind ejecta.

To explore this effect, we created several models in which we
superimposed a  spherical shell of $\Ye < 0.25$ material
onto the outer region of the 2D wind model t100.  We took the radial
density profile of the shell to be a gaussian with central velocity $v
= 0.2c$ and a width of $\Delta v = 0.1c$ and varied the mass of
dynamical ejecta between $M_{\rm dyn} = 10^{-5}$ and
$10^{-2}~M_\odot$.

\begin{figure*}
\includegraphics*[width=6.8in]{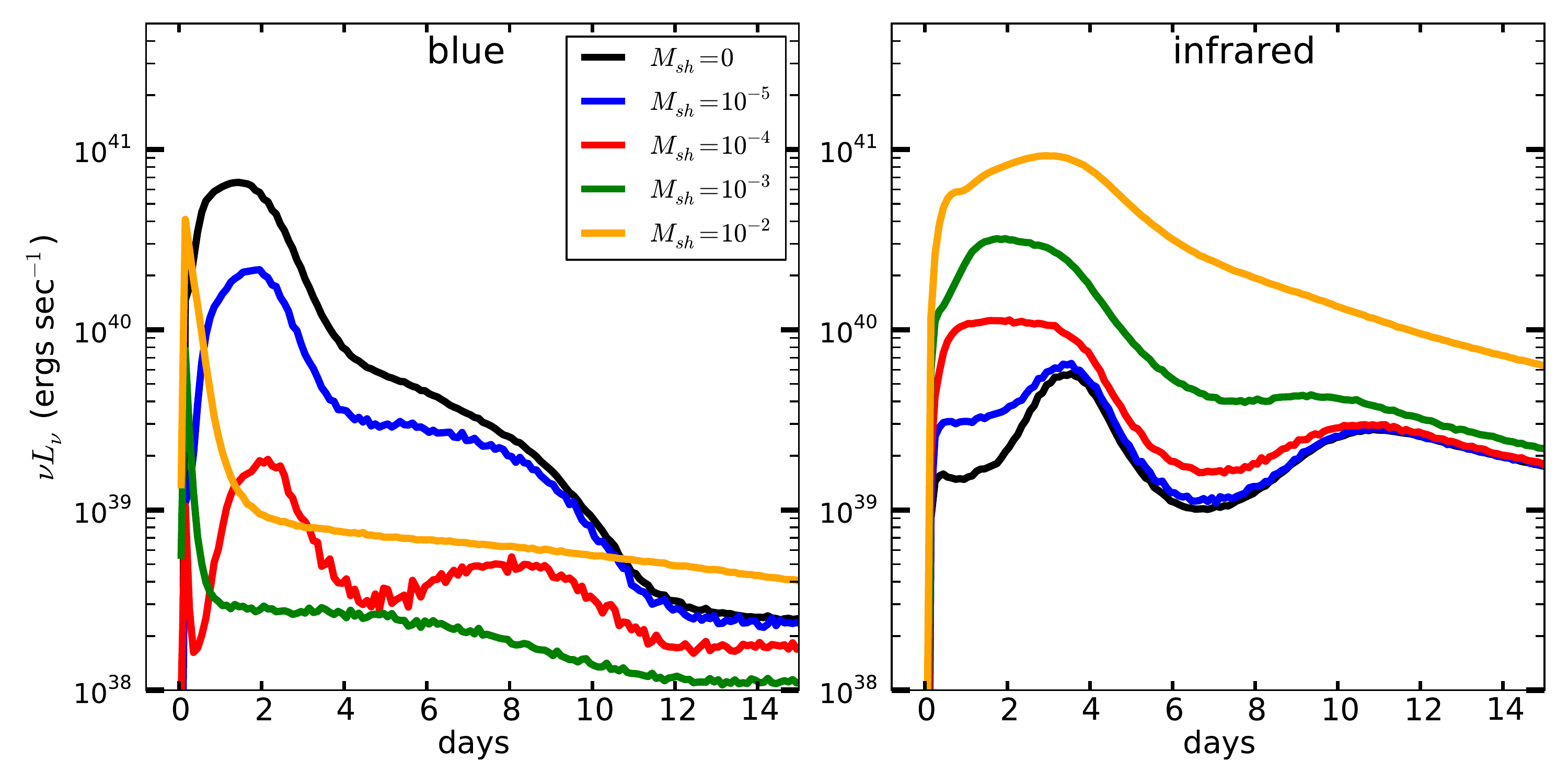}
\caption{
 Angle averaged model light curves at  optical blue wavelengths (left panel) and infrared wavelengths (right panel) for model
 t100 surrounded by a varying amount of spherically distributed dynamical ejecta.  Only a small amount ($\sim 10^{-4} M_\odot$) of low-\Ye\ dynamical ejecta is needed to obscure  the optical emission from the wind.
 \label{fig:spheres}
}
\end{figure*}

Figure~\ref{fig:spheres} shows the angle-averaged light curves for
models with different values of $M_{\rm dyn}$.  Due to the extremely
high lanthanide opacities, we find that
only a small mass of dynamical ejecta ($M_{\rm dyn} \approx
10^{-4}~M_\odot$) is required to suppress the blue optical flux by more
than an order of magnitude.  For models with $M_{\rm dyn} \gtrsim
10^{-3}~M_\odot$, the wind emission is completely invisible, and the optical
and infrared light curves arise entirely  within  the
dynamical ejecta.  In these cases, the brief peak and subsequent tail
of blue flux are due to the small fraction of optical radiation that
is produced in and manages to escape from the heavily line-blanketed
dynamical ejecta.

In realistic merger simulations, the dynamical ejecta is not
completely spherical, and may possess some lanthanide-free ``windows"
through which we can see the wind ejecta.  In the NS + NS simulations of
\cite{Bauswein+13}, the outflows have a torus-like anisotropy, with
less material ejected along the polar axis.  Neutrino irradiation of
the dynamical ejecta may also raise the \Ye\ in the polar region,
reducing the abundance of lanthanides in that material
\citep{Wanajo+14}.  In BH-NS mergers, mass ejection occurs mainly
through the tidal tail of the NS, and forms a single thick arm
confined to the equatorial plane.

\begin{figure}
\includegraphics*[width=\columnwidth]{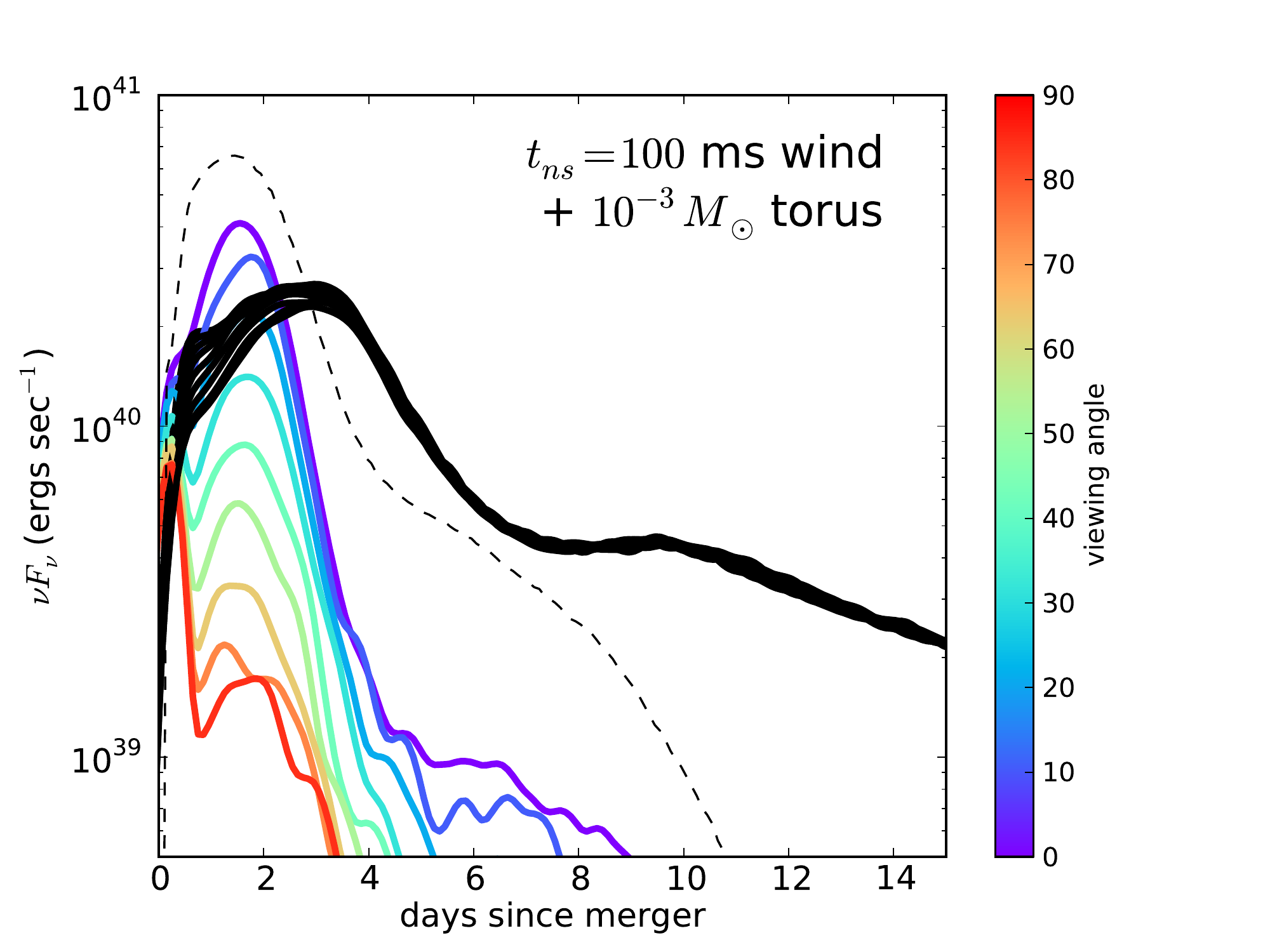}
\caption{Synthetic light curves of 
wind model t100 surrounded by a $10^{-3}~M_\odot$ torus 
of heavy r-process elements.  Colored lines denote the optical blue  
($3500-5000$~\AA) light curve as seen from different viewing angles, while  solid black lines denote the infrared light curve.  The dashed black line shows the angle averaged blue light curve of model t100 when no dynamical ejecta is included.  For pole on viewing angles ($\theta \approx 0^\circ$) the blue emission is visible through the hole in the torus, but for edge on 
orientations ($\theta \approx 90^\circ$), the  dynamical ejecta  suppresses the optical flux. \label{fig:torus_wind}
}
\end{figure}

To explore the possible geometrical effects, we constructed an
additional model in which $M_{\rm dyn} = 10^{-3}$ of dynamical ejecta
was distributed in a torus about the equatorial plane.  Here we took
the density profile in both the $r$ and $z$ directions to be gaussian
with width of $\Delta v = 0.1c$, a distribution that closely resembles
the ejecta seen in the Newtonian NS-NS simulations of
\cite{Rosswog05}. Figure~\ref{fig:torus_wind} shows the resulting
light curves as seen from different viewing angles.  For pole-on
orientations ($\theta = 0^\circ$), an observer can see most of the
wind ejecta through the hole of the dynamical torus, and the blue
optical flux is only reduced by factor of $\sim 2$ at peak.  For
orientations closer to edge-on ($\theta = 90^\circ$), however, the
dynamical torus obscures the wind optical flux by an order of
magnitude or more.

\subsection{Comparison to Observations}
\label{sec:obs}

One week following the gamma-ray burst (GRB) 130603B, \cite{Tanvir+13}
and \cite{Berger+13} detected an excess infrared emission that they
attributed to a kilonova.  In the right panel of Figure
\ref{fig:allmods}, we compare that {\it Hubble Space Telescope}
infrared data point to our models.  The luminosity of the 130603B
excess is a factor of $\sim 5$ higher than the peak of even our
brightest disk wind models.  Since the peak luminosity is proportional
to the square root of the ejecta mass, fitting the GRB 130603B
infrared excess appears to require an ejecta mass of $\sim
0.1~M_\odot$, a factor of $\sim 20$ times higher than that ejected in any
of our models.

Disk winds could explain the large ejecta mass inferred for
GRB 130603B if we assume that the accretion disk was of higher initial
mass, say $\sim 0.1- 0.3M_{\odot}$, rather the value $0.03M_{\odot}$
adopted in our calculations.  In addition, most of our models assumed
a non-spinning black hole, whereas a rapidly spinning black hole can
increase the wind ejecta mass by a factor of $\sim 5$ \citep{Just+14,
  Fernandez+14}.  Dynamical ejecta may also contribute to the infrared
emission.  A mass of $\sim 0.1M_{\odot}$ is likely too large to come
solely from the dynamical ejecta of a NS-NS merger, but it might be
expelled in the merger between a NS and a low mass BH
(\citealt{Tanaka+14}).

The left panel of Figure \ref{fig:allmods} compares our blue optical
light curves to a different possible kilonova candidate, the optical
``bump" that followed GRB 080503 (\citealt{Perley+09}).  This event
possessed no clear host galaxy and so we have chosen a redshift $z =
0.25$ to place the observations at a similar brightness to our models.
The light curves of disk wind model with a long lived HMNS (e.g.,
tInf) fit the data reasonably well.  A wind model that ejected $\sim
4$ times more mass would likely better fit the slower observed light
curve decline, and would place the kilonova at the redshift $z =
0.56$, which corresponds to a faint spiral galaxy within the field of
080503 (a strong natal kick of the binary would be required to explain
its large observed spatial offset in this case).  GRB 080503 was
accompanied by extremely bright extended prompt X-ray emission, which
\citet{Metzger+08b} speculate is powered by a stable magnetar created
during a NS-NS merger (see also \citealt{Bucciantini+12},
\citealt{Rezzolla&Kumar14}).  In this case, the kilonova emission
could be significantly enhanced by rotational energy injected by the
magnetized remnant (\citealt{Yu+13}; \citealt{Metzger&Piro13}).


\section{Summary and Conclusions}
\label{sec:conclude}

\begin{figure*}
\includegraphics*[width=7in]{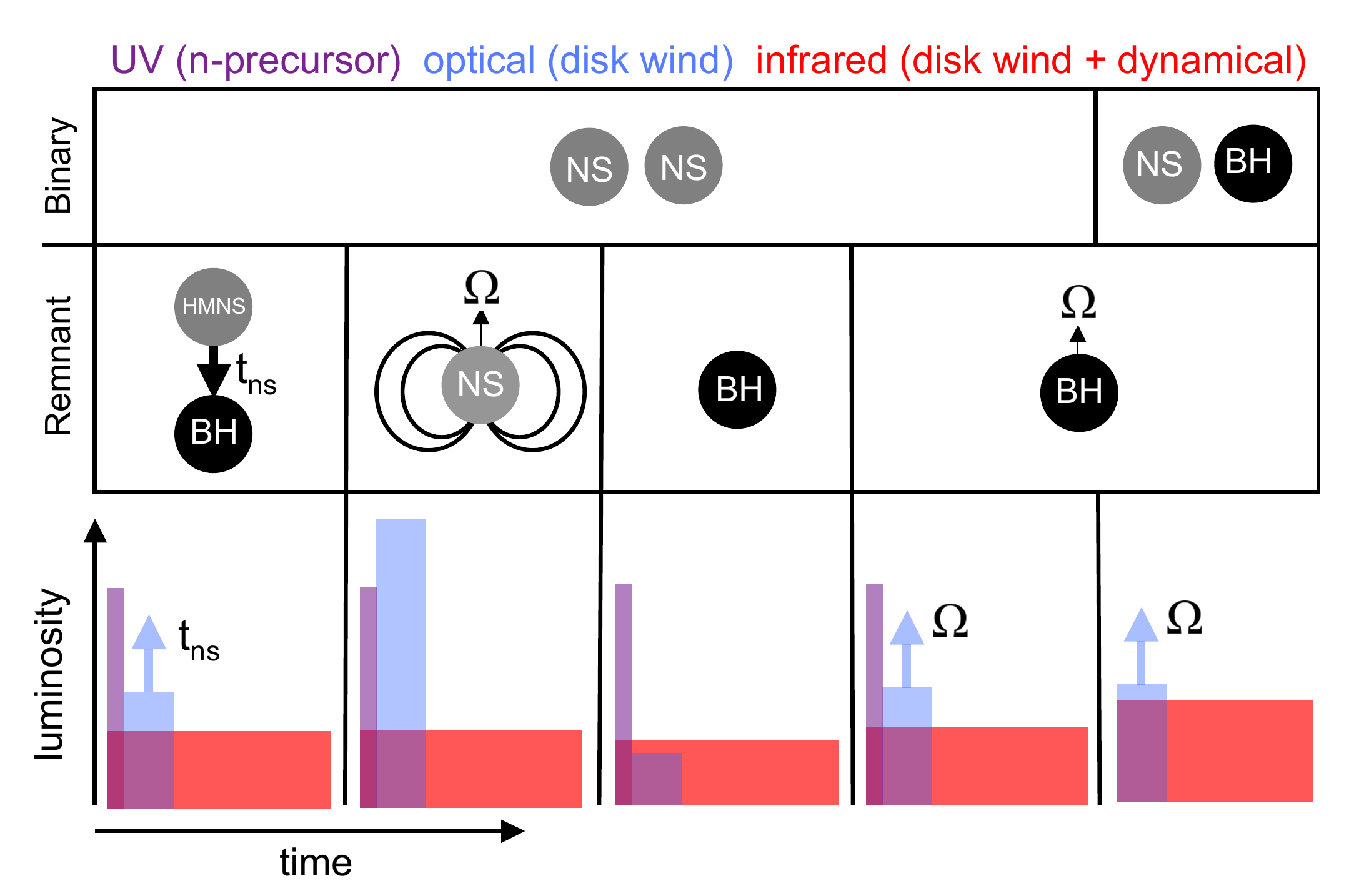}
\caption{ 
Schematic illustration of the mapping between mergers and kilonova light curves. 
 The top panel shows the progenitor system, either a NS-NS or a NS-BS binary, while the
the middle plane shows the final merger remnant (from left to right: a HMNS that collapses to a BH after time \tns, a
spinning magnetized NS, a non-spinning BH, and a rapidly spinning BH).  The bottom panel illustrates the relative amount of
UV/blue emission from a n-precursor (purple),  optical emission from lanthanide-free material (blue), and IR emission from lanthanide containing ejecta (red).
\label{fig:schematic} }
\end{figure*}

We have shown how material ejected in disk winds  subsequent to a compact object merger
can give rise to both optical and infrared kilonova emission. 
We studied the dependences  on key parameters
such as the delay until black hole formation, black hole spin,
and the presence of neutron-rich dynamical ejecta. 
In Figure 9, we  summarize schematically the range of possible kilonova properties, and
illustrate how they roughly map to the progenitor binary and remnant type.  Our main findings are as follows: 
\newline

\noindent 1. -- For the characteristic entropies and expansion times
of disk winds, we find that the abundance of lanthanides cuts off
sharply when the electron fraction $Y_e \gtrsim 0.25$
(Fig.~\ref{fig:torch}).  The electron fraction in turn is very
sensitive to the level of neutrino irradiation of the wind and hence
acts as a diagnostic of the physical conditions in the aftermath of
the merger (\citealt{Metzger&Fernandez14}).
      \newline

\noindent 2. -- The presence of  optical emission  is a
ubiquitous feature of the disk wind ejecta, even in the case of
non-spinning, promptly-formed black hole remnants
(Fig.~\ref{fig:br_lc}).  The magnitude and duration of this optical component
is a sensitive function of the lifetime of a HMNS or the spin of the
promptly-formed BH. In the limit of a very long-lived HMNS, 
photons emerge primarily in the optical, reaching
luminosities up to $10^{41}$~erg~s$^{-1}$ for moderate disk masses
($0.03M_\sun$).
      \newline

\noindent 3. -- The ratio of the optical to infrared luminosity from a
kilonova provides a powerful measure of the relative mass of high-\Ye\ to
low-\Ye\ ejecta.  Using this information to infer the underlying the
physical scenario, however, may be difficult given the degeneracies.
For example, the wind ejecta from a promptly-formed, rapidly spinning
BH produces a similar kilonova light curve to that of a long-lived
HMNS (Figure~\ref{fig:allmods}).
\newline

\noindent 4. -- Because the expansion velocities of the wind are
moderate ($\sim 10,000$~km~s$^{-1}$), numerous line absorption
features are discernible in the spectra
(Fig.~\ref{fig:spec_series}).  This distinguishes the spectra of disk winds from
those of fast moving dynamical ejecta, for which the line features are broader
and heavily blended.  Observing the spectra of wind ejecta may thus
allow us to study the detailed composition of freshly produced
r-process material.  At present, however, the atomic data is not good
enough to predict all line wavelengths, and additional atomic
structure calculations are needed.
      \newline

\noindent 5. -- The optical emission from a disk wind can be easily
obscured by even a small amount ($10^{-4}M_\sun$) of neutron-rich
dynamical ejecta, causing most of the flux to emerge in the near
infrared.  In the case of NS-NS mergers, this dynamical ejecta is
expected to be nearly isotropic, reducing the likelihood of observing
an optical component (Fig.~\ref{fig:spheres}). For BH-NS mergers, the
confinement of the dynamical ejecta to the equatorial plane makes the
detection of optical emission possible from polar viewing angles
(Fig.~\ref{fig:torus_wind}).
      \newline

\noindent 6. -- The infrared emission excess observed following
GRB~130603B can be explained by a disk wind only if the merger formed
a rather massive disk ($\sim 0.1~M_\odot$).  The optical bump observed
following GRB~080503 (with a redshift of $z = 0.25 - 0.5$) can be nicely
explained by a wind from a moderately massive disk ($\sim
0.03~M_\odot$) irradiated by a long-lived HMNS.
\newline

Our calculations have illustrated the key properties of kilonova from disk winds, however several
improvements are needed to generate reliable, quantitative theoretical
predictions.  On the wind dynamics side, a more advanced neutrino
transport scheme is required to better quantify the distribution of
electron fraction in the ejecta.  Inclusion of magneto-hydrodynamics
and general relativity is also important to quantify the amount of 
ejected mass.  Using more realistic initial conditions,
taken from an actual merger simulation, would provide a better description of
the relative size and spatial distribution of the dynamical ejecta and
accretion disk.  Determining the detailed final composition of the
wind requires post-processing wind tracer particles with nuclear
reaction networks.  On the radiative transport side, better line data
for the high-Z elements, in particular the lanthanides and actinides,
is necessary to calculate the pseudo-continuum opacity and line
features.  We also need a better understanding of the radioactive
decay energy rate, and the thermalization of decay products
(electrons, gamma-rays, and fission fragments).  Our future work will
take steps along these lines, with the hope of developing more
realistic kilonova models.


\section*{Acknowledgments}
DK is supported in part by a Department of Energy Office of Nuclear
Physics Early Career Award, and by the Director, Office of Energy
Research, Office of High Energy and Nuclear Physics, Divisions of
Nuclear Physics, of the U.S. Department of Energy under Contract No.
DE-AC02-05CH11231. 
RF was supported 
by NSF Division of Astronomical Sciences collaborative research grant AST-1206097,
and a UC Office of the President grant.
BDM gratefully acknowledges support from the NSF grant AST-1410950 and the Alfred P. Sloan Foundation.
This work was supported in part by NSF Grant No. PHYS-1066293 and the hospitality of the Aspen Center for Physics.
The software used in this work was in part developed by the DOE NNSA-ASC OASCR Flash Center at the
University of Chicago.
We are grateful for computing time made available by
the National Energy Research Scientific Computing Center, which is supported by the Office of 
Science of the U.S. Department of Energy under Contract No. DE-AC02-05CH11231.  
Computations were performed using  \emph{Carver} and \emph{Hopper}.


\bibliographystyle{mn2e}
\bibliography{ms}

\label{lastpage}

\newpage

\clearpage

\end{document}